# Well Placement Optimization under Uncertainty with CMA-ES Using the Neighborhood

*Z. Bouzarkouna\* (TOTAL (Formerly IFP Energies nouvelles)), D.Y. Ding (IFP Energies nouvelles) & A. Auger (INRIA)*

In the well placement problem, as well as in many other field development optimization problems, geological uncertainty is a key source of risk affecting the viability of field development projects. Well placement problems under geological uncertainty are formulated as optimization problems in which the objective function is evaluated using a reservoir simulator on a number of possible geological realizations. The existing approaches to cope with geological uncertainty require multiple reservoir simulations (on the possible realizations) to estimate the expected field performance for a given well configuration, which is often very demanding in CPU time and impractical when dealing with real field cases.

In this paper, we present a new approach to handle geological uncertainty for the well placement problem with a reduced number of reservoir simulations. The proposed approach uses already simulated well configurations in the neighborhood of each well configuration for the objective function evaluation. We use thus only one single reservoir simulation performed on a randomly chosen realization together with the neighborhood of each well configuration to estimate the objective function instead of using multiple simulations on multiple realizations. This approach is combined with the stochastic optimizer CMA-ES (Covariance Matrix Adaptation - Evolution Strategy).

The proposed approach is shown on the benchmark reservoir case PUNQ-S3 to be able to capture the geological uncertainty using a smaller number of reservoir simulations. This approach is compared to the reference approach using all the possible realizations for each well configuration. It is shown that the proposed approach is able to reduce significantly the number of reservoir simulations by more than 80% for the reservoir case in this study.

---

\* Now with TOTAL.

## 1. Introduction

Many of the oil and gas field development problems can be formulated as optimization problems. Ranging from production optimization to well placement optimization, the use of optimization methods becomes more and more popular. The objective function optimized, which is usually evaluated using a reservoir simulator, is defined in a way to evaluate the economics of the project (i.e., the NPV). It can also simply be defined as the cumulative oil production. The parameters of the problem encode the decision variables to be optimized (e.g., the position of the different wells in well placement optimization, or the bottom-hole pressures or the flow-rates in production optimization).

Many optimization methodologies were used in the literature to tackle field development problems. Approaches based on stochastic search algorithms were used such as genetic algorithms (Guyaguler & Horne, 2000; Guyaguler et al., 2000; Yeten et al., 2003), simulated annealing (Beckner & Song, 1995; Norrena & Deutsch, 2002), particle swarm optimization (Onwunalu & Durlofsky, 2010), CMA-ES (Bouzarkouna et al., 2011; Bouzarkouna et al., 2012). Deterministic optimization methods were also used in some studies such as in Handels et al. (2007), Sarma & Chen (2008), Forouzanfar et al. (2010).

However, in the well placement problem, as well as in many other field development optimization problems, geological uncertainty is a key source of risk affecting the viability of field development projects, although still neglected in many research studies. The problem arises when we have multiple possible geological realizations of the reservoir. The multiple realizations are generated using geostatistical techniques and in general deemed equiprobable. Taking into account these several geological realizations adds an important challenge to the optimization of the objective function; in particular it leads to a large increase of the number of performed reservoir simulations (Usually, a single objective function evaluation requires a number of reservoir simulations equal to the number of considered realizations).

In this paper, we propose a new approach to handle geological uncertainty with a reduced number of reservoir simulations. The proposed approach is demonstrated on the well placement problem, but can be extended to other Geosciences problems such as production optimization.
In particular, this paper addresses the problem of how to define the objective function when dealing with uncertainty for well placement and whether we should perform evaluations on all the possible realizations in order to define the objective function.
Thus, we propose to use only a small number of reservoir simulations (typically one) for each well configuration, together with its neighborhood in order to estimate its corresponding objective function value instead of using multiple realizations.

We denote by $f : \Re^n \to \Re$ the objective function to optimize and by $N_r$ the number of geological realizations which are denoted by $(R_i)_{i=1,\dots,N_r}$. For each well configuration proposed by the optimizer, we have $N_r$ possible values of the objective function, one for each realization where each will be denoted for a given well configuration $\mathbf{x}$ by $f(\mathbf{x}, R_i)$ corresponding to a the realization $R_i$.

This paper is structured as follows. Section 2 provides a detailed literature review for well placement optimization under geological uncertainty. Section 3 defines a new approach to handle geological uncertainty for well placement using the neighborhood, which is combined with the optimization method CMA-ES. In Section 4, we demonstrate the contribution of the proposed approach in capturing the geological uncertainty and in reducing the number of reservoir simulations on the synthetic benchmark reservoir case PUNQ-S3 (Floris et al., 2001).

## 2. Optimization under uncertainty: a literature review

In the context of field development optimization under geological uncertainty, the measurable fitness values correspond to the values $f(\mathbf{x}, R_i)_{i=1,\ldots,N_r}$ which represent the evaluations on each of the realizations. Therefore, the objective function can be in general written as:

$$f(\mathbf{x}) = \frac{1}{N_r} \sum_{i=1}^{N_r} \left[ f(\mathbf{x}, R_i) \right] . \tag{1}$$

However due to the expensive computational effort required to evaluate the objective function over one realization $R_i$, the expected objective function is in some cases approximated in a way to use a fewer number of samples instead of using all the realizations. Thus, one common way to approximate the expected objective function here is by averaging over a number of samples $N_s \leq N_r$.

The problem of optimization under geological uncertainty shares many similarities with the problem of optimizing noisy functions. A function $f$ is said to be noisy if the only measurable value of $f$ on $\mathbf{x} \in \Re^n$ is a random variable that can be written as $F(f(\mathbf{x}), z)$ where f is a time-invariant function and $z$ is a noise often assumed to be normally distributed with a zero mean and variance $\sigma^2$, and denoted by $\aleph(0, \sigma^2)$. The noise can be also defined differently (e.g., Cauchy distributed), and can be either additive or multiplicative. To optimize noisy functions, the objective function is usually estimated by the expected value defined as follows:

$$f(\mathbf{x}) = \int_{-\infty}^{+\infty} \left[ F(f(\mathbf{x}), z) \right] p(z) dz , \tag{2}$$

where $p(z)$ is the probability density function of the noise. Thus, a common way to approximate the expected objective function is again by averaging over a finite number of random samples $N_s$:

$$f(\mathbf{x}) = \frac{1}{N_s} \sum_{i=1}^{N_s} \left[ F(f(\mathbf{x}), z_i) \right] , \tag{3}$$

In the following, we briefly review the existing approaches often used in optimization under uncertainty. On the one hand we review the approaches defined by the optimization community mainly to cope with noise but that can be extended to the different field development optimization under geological uncertainty. On the other hand we review the approaches already applied in the petroleum community to cope with geological uncertainty.

### 2.1. Optimization community

This section summarizes the different ways to handle uncertainty within the evolutionary optimization community. A detailed overview of the existing approaches addressing uncertainties in evolutionary optimization is presented by Jin & Branke (2005). Let us suppose in this section that the function $f$ to optimize is a noisy function. The approaches to handle uncertainty are mainly divided into two categories.

### 2.1.1. Explicit Averaging

*Using mean of several samples for each individual*
The simplest and the most common way to address the uncertainty issue is to define the objective function for each point by averaging over a number of samples (Eq. (3)). Increasing the sample size $N_s$ is equivalent to reducing the variance of estimating the objective function. In general, the objective function is defined using an averaged sum of a *constant sample size*. In this case, for each single evaluation of the expected objective function, one needs to evaluate the objective function on $N_s$ samples. In the context of costly objective functions, depending on the number of samples, there is a compromise between the computational cost of the optimization and the accuracy of the estimation of the objective function. Increasing (respectively, decreasing) the number of samples tends to improve (respectively, worsen) the accuracy of the estimated objective function, but on the other hand it tends also to increase (respectively, reduce) the computational cost of the optimization. The idea of using an *adapted sample size* during the optimization was first proposed by Aizawa & Wah (1993) and Aizawa & Wah (1994). In Aizawa & Wah (1994), it is shown that adapting the number of samples performs better than using constant sample sizes, and it is suggested to increase the sample size with the generation number and to use a higher number of samples for individuals with higher estimated variance. Another way to adapt the sample size is based on an individual's probability to be among a number of the best individuals (Stagge, 1998). Recently, another approach relying on the rank based selection operators was proposed by Hansen et al. (2009). In Heidrich-Meisner & Igel (2009), an adaptive uncertainty handling procedure is proposed, based on selection races.

*Using the neighborhood for each individual*
An alternative approach to defining the objective function as an averaged sum of a number of samples (constant or adapted) is to define the objective function using the neighborhood points already evaluated (Pänke et al, 2006; Branke et al., 2001; Branke & Schmidt, 2005; Sano & Kita, 2000; Sano & Kita, 2002). The general idea has first been suggested by Branke (1998) in which it is suggested to estimate the fitness as a weighted average of the neighborhood with a linearly decreasing weight function up to some fixed maximum distance. In Pänke et al (2006), Branke et al. (2001) and Branke & Schmidt (2005), a locally weighted regression is used for estimation. This technique is shown to be a good solution to improve the accuracy of the estimated objective function without increasing the computational cost.

### 2.1.2. Implicit Averaging

When increasing the population size, the probability to obtain similar points is higher. Thus, a way to cope with noise is to simply increase the population size (Fitzpatrick & Grefenstette, 1988). In this case, with a large population size, the influence of noise on a given point can be reduced due to the evaluations on other similar points. Conflicting conclusions (Fitzpatrick & Grefenstette, 1988, Arnold & Beyer, 2000; Arnold & Beyer, 2001; Beyer, 1993) were shown in the literature when comparing explicit averaging and implicit averaging.

### 2.2. Petroleum community

Several studies in the literature have addressed the problem of optimization under geological uncertainty. Optimization under geological uncertainty in the petroleum community considers always a finite number of realizations $N_r$ and models the objective function following Eq. (1). In the following we briefly review the approaches to handle uncertainty in optimization within the petroleum community.

To the best of our knowledge, all the studies that consider a number $N_r$ of possible geological realizations use the approach "Using mean of several samples for each individual". Moreover, all the studies in the literature, except the approach proposed in Wang et al. (2012) that will be detailed later, perform $N_r$ reservoir simulations for every single objective function evaluation. Although sharing this common similarity, the proposed approaches introduce different formulations of the objective function.

In Schulze-Riegert et al. (2010), Schulze-Riegert et al. (2011), Onwunalu & Durlofsky (2010) and Chen (2010), the objective function is formulated as the expected value of the net present value over all the realizations, as shown in Eq. (1). Chen (2010) tackles the problem of closed-loop production optimization using the optimizer EnOpt (Chen & Oliver, 2009; Chen et al., 2009) which is applied to the geological model ensemble updated by either EnKF (Evensen, 1994) or EnRML (Gu & Oliver, 2007).

In Yeten et al. (2003), Aitokhuehi et al. (2004) and Alhuthali et al. (2010), multiple geostatistical realizations of the reservoir are considered in the formulation of the objective function:

$$f(\mathbf{x}) = \frac{1}{N_r} \sum_{i=1}^{N_r} \left[ f(\mathbf{x}, R_i) \right] + r\sigma \quad , \tag{4}$$

where $r \in \Re$ is the risk factor and $\sigma$ is the standard deviation of $f$ on $\mathbf{x}$ over the realizations, defined as follows:

$$\sigma = \sqrt{\frac{1}{N_r} \sum_{i=1}^{N_r} \left( f(\mathbf{x}, R_i) - \langle f(\mathbf{x}) \rangle \right)^2} \quad , \tag{5}$$

where:

$$\langle f(\mathbf{x}) \rangle = \frac{1}{N_r} \sum_{i=1}^{N_r} \left[ f(\mathbf{x}, R_i) \right] \quad . \tag{6}$$

The term $r\sigma$ in Eq. (4) is used to take into account the decision maker's attitude toward risk. A positive $r$ indicates a risk-prone attitude, a negative $r$ indicates a risk-averse attitude and an $r = 0$ indicates a risk-neutral attitude. This formulation is close to the formulations defined in Guyaguler & Horne (2001) and Ozdogan & Horne (2006) using utility functions.

In Artus et al. (2006), a more general formulation of the objective function is defined as follows. A genetic algorithm is used, in which at each iteration only a predefined percentage of the individuals, chosen according to a set of scenario attributes, is simulated. For the simulated individuals, Artus et al. propose to perform again $N_r$ reservoir simulations for each well configuration $\mathbf{x}$ in order to evaluate the values of $f(\mathbf{x}, R_i)$ on all realizations and then to derive the cumulative distribution function $\text{cdf}\{f\}$ on $\mathbf{x}$. From this distribution, the values of $f^{10}(\mathbf{x})$, $f^{50}(\mathbf{x})$ and $f^{90}(\mathbf{x})$ are determined. The value $f^{10}$ on $\mathbf{x}$ denotes the value of $f$ on $\mathbf{x}$ corresponding to a probability of 0.1, i.e., there is a probability 0.1 that the value of $f$ on $\mathbf{x}$ will be less than $f^{10}$ on $\mathbf{x}$. The value $f^{10}$ on $\mathbf{x}$ can be written as $\text{cdf}\{f\}^{-1}(0.1)$. The values $f^{50}(\mathbf{x})$ and $f^{90}(\mathbf{x})$ are defined in a way similar to $f^{10}(\mathbf{x})$. The objective function is then formulated as follows:

$$f(\mathbf{x}) = r_{10} f^{10}(\mathbf{x}) r_{50} f^{50}(\mathbf{x}) + r_{90} f^{90}(\mathbf{x}) \quad , \tag{7}$$

where the parameters $r_{10}$, $r_{50}$ and $r_{90}$ are defined according to the decision maker's attitude toward risk. A risk-neutral attitude corresponds to the case where $(r_{10}, r_{50}, r_{90}) = (0, 1, 0)$. However, a risk-averse investor tends to increase the value of $r_{10}$, and a risk-prone investor tends to increase the value of $r_{90}$.

Another way to formulate the objective function under geological uncertainty is to optimize the worst case scenario using a min-max problem formulation. This approach is used in Alhuthali et al. (2010) to optimize smart well controls.

The only approach in the literature selecting only a number of samples instead of all the realizations is defined in Wang et al. (2012). The approach is based on the so-called retrospective optimization and divides the problem into a number of sub-problems, where the initial solution of the current sub-problem is simply the returned solution from the previous sub-problem. Each point to be evaluated is approximated by the average over a number of realizations, where the number of selected realizations is increased from sub-problem to sub-problem. The approach implies then defining a sequence of samples. The example shown in Wang et al. (2012) considers a well placement problem on 104 permeability and porosity realizations and therefore defines sub-problems with a sequence {20,15,10,5} of iterations and a sequence {1,5,16,21,104} of sample sizes. Although Wang et al. suggest further testing of the overall framework to determine the appropriate sequence of sample sizes, an answer can be the work on adapting automatically the sample sizes already proposed in Stagge (1998) or in Hansen et al. (2009) but still demanding in the number of objective function evaluations.

## 3. Well placement under uncertainty with CMA-ES using the neighborhood

This section proposes a new approach to handle geological uncertainty for well placement. The proposed approach focuses on reducing the uncertainty by using the objective function evaluations of already evaluated individuals in the neighborhood. In this section, we propose then to apply an approach based on using the neighborhood for each individual. The approach is based on the optimization algorithm Covariance Matrix Adaptation – Evolution Strategy (CMA-ES), but can be combined with any other optimization algorithm.

*The optimization algorithm: CMA-ES*

In the following, we describe the CMA-ES (Hansen and Ostermeier, 2001). CMA-ES is a stochastic population-based optimization algorithm where at each iteration step $g$, a population of $\lambda$ points is generated by sampling a multivariate normal distribution. The objective function values of the $\lambda$ points are then evaluated and the parameters of the multivariate normal distribution are updated. More specifically, let $(\mathbf{m}^g, g \in \aleph)$ be the sequence of mean values of the multivariate normal distribution and let $(\sigma^g, g \in \aleph)$ and $(\mathbf{C}^g, g \in \aleph)$ be respectively the sequences of step-sizes and covariance matrices. The sampling of the $\lambda$ points of the new population ($\mathbf{x}_i^g)_{i=1,\dots,\lambda}$ can then be written as:

$$\mathbf{x}_i^g = \mathbf{m}^g + \sigma^g N_i(0, \mathbf{C}^g), \quad \text{for } i = 1, \dots, \lambda \ , \tag{8}$$

where $N_i(0, \mathbf{C}^g)$ are independent multivariate normal distributions with zero mean vector and covariance matrix $\mathbf{C}^g$. A more detailed overview of CMA-ES and its application on the well placement problem can be found in Bouzarkouna et al. (2011).

*Using the neighborhood approach*

We define a CMA-ES optimizing an estimated fitness defined on a given point using a weighted average of a small number of evaluations on the considered point and a number of evaluations already performed on the neighborhood (up to some fixed maximum distance) with a decreasing weight function depending on the Mahalanobis distance –between the considered point and the neighbor point– with respect to the covariance matrix $\mathbf{C}$ defined by CMA-ES. Although considering a Mahalanobis distance with respect to $\sigma^2 \mathbf{C}$ is suspected to be a better choice (since we are using a fixed maximum distance to select the neighbors), it has not been tested in this paper.

Let us consider a well placement optimization problem with a number of wells (producers and/or injectors) to be placed. Let us denote by $n$ the dimension of the problem, i.e., the number of parameters needed to encode the wells to be placed. The wells to be placed are parameterized using real numbers encoding its coordinates.

Without loss of generality, we will consider in the sequel the NPV as the objective function that we aim to optimize, unless otherwise explicitly stated. Thus, we want to find a vector of parameter $\mathbf{p}_{\max,R} \in \mathfrak{R}^n$ such that:

$$\text{NPV}^R(\mathbf{p}_{\max,R}) = \max_{\mathbf{p}} \left\{ \text{NPV}^R(\mathbf{p}) \right\} \ , \tag{9}$$

where $\text{NPV}^R$ is the averaged sum of the NPVs of a given well configuration represented by a vector of parameters $\mathbf{p}$ over all the realizations:

$$\text{NPV}^R(\mathbf{p}) = \frac{1}{N_r} \sum_{i=1}^{N_r} \text{NPV}^R(\mathbf{p}, R_i) \ . \tag{10}$$

In the proposed approach, we define a so-called estimated objective function that will be optimized instead of the true objective function $\text{NPV}^R$ defined in Eq (10). The estimated function will be denoted in the sequel by $\text{NPV}^E$. Thus in the proposed approach, contrary to what is shown in Eq (9), we will try to find the vector of parameter $\mathbf{p}_{\max,E} \in \mathfrak{R}^n$ such that:

$$\text{NPV}^E(\mathbf{p}_{\max,E}) = \max_{\mathbf{p}} \left\{ \text{NPV}^E(\mathbf{p}) \right\} \ . \tag{11}$$

The simplest case, in which solving Eq. (9) is equivalent to solving Eq. (11), is when $\text{NPV}^E$ is a monotonic transformation of $\text{NPV}^R$. However in this work, we do not aim to define an estimated objective function $\text{NPV}^E$ such that we can prove that Eq. (11) is equivalent to Eq. (9). Our aim is that by solving Eq. (11), we can propose good points with high $\text{NPV}^R$ values (see below for the definition of $\text{NPV}^E$).

To optimize $NPV^E$, we propose to use the CMA-ES optimizer. During the optimization process, we build a database -called also training set- in which after every performed reservoir simulation for a given point $\mathbf{x}$ on a realization $R$, we store the point $\mathbf{x}$ together with its corresponding evaluation $NPV(\mathbf{x}, R)$.

It remains now to define the estimated objective function $NPV^E$ for a given point (well configuration) denoted by a vector of parameters $\mathbf{p}$ :

1. At the beginning of the optimization and until reaching a given number $N_{sim}$ of performed reservoir simulations, we define a number of reservoir simulations $N_s^1$ ($\leq N_r$) to be performed on $\mathbf{p}$, and a set of $N_s^1$ randomly drawn integers $\{j_1,...,j_{N_s^1}\} \subset \{1,...,N_r\}$. We perform then $N_s^1$ reservoir simulations on $\mathbf{p}$ on the realizations $(R_i)_{i=j_1,...,j_{N_s^1}}$, and we add each of the obtained simulation results $(\mathbf{p}, NPV(\mathbf{p}, R_i))$ to the training set.

   The estimated objective function on the point $\mathbf{p}$ reads as follows:

   $$NPV^E(\mathbf{p}) = \frac{1}{N_s^1}\sum_{i=1}^{N_s^1} NPV(\mathbf{p}, R_{j_i}) \ . \tag{12}$$

   In this case, the evaluation of $NPV^E$ requires a number $N_s^1$ of reservoir simulations.

2. If more than $N_{sim}$ reservoir simulations are performed, we perform the following steps.

   We begin by defining a number of reservoir simulations $N_s^2$ ($\leq N_r$) to be performed on $\mathbf{p}$, and a set of randomly drawn integers $\{j_1,...,j_{N_s^2}\} \subset \{1,...,N_r\}$. We perform then $N_s^2$ reservoir simulations on $\mathbf{p}$ on the realizations $(R_i)_{i=j_1,...,j_{N_s^2}}$, and we add each of the obtained simulation results $(\mathbf{p}, NPV(\mathbf{p}, R_i))$ to the training set.

   We also define a maximum number of neighbor points $N_{n,max} \in \mathbb{N}$ that can be used in the definition of $NPV^E$. We select then at most the $N_{n,max}$ nearest points to $\mathbf{p}$ from the training set. Here, we select only the points with a distance less or equal to a given fixed distance of selection denoted by $d_{max}$. We denote by $N_n$ the number of selected points and by $(\mathbf{x}_i)_{1\leq i\leq N_n}$ the selected points[1]. The distance used for this purpose is the Mahalanobis distance with respect to the current covariance matrix $\mathbf{C}$ of CMA-ES defined for two given points $\mathbf{z}_1 \in \mathfrak{R}^n$ and $\mathbf{z}_2 \in \mathfrak{R}^n$ by $d_{\mathbf{C}}(\mathbf{z}_1, \mathbf{z}_2) = \sqrt{(\mathbf{z}_1 - \mathbf{z}_2)^T \mathbf{C}^{-1}(\mathbf{z}_1 - \mathbf{z}_2)}$ .

   The estimated objective function on $\mathbf{p}$ reads as follows:

   $$NPV^E(\mathbf{p}) = \frac{1}{S}\left[\sum_{i=1}^{N_s^2}\left(p_i NPV(\mathbf{p}, R_{j_i})\right) + \sum_{i=1}^{N_n}\left(\widetilde{p}_i NPV(\mathbf{x}_i, R_i)\right)\right] \ , \tag{13}$$

   where $p_i = 1$, $\widetilde{p}_i = \left(1 - \left(\frac{d_{\mathbf{C}}(\mathbf{x}, \mathbf{p})}{d_{max}}\right)^2\right)^2$ and $S = \sum_{i=1}^{N_s^2} p_i + \sum_{i=1}^{N_n}\widetilde{p}_i$ . In this case, the evaluation of $NPV^E$ requires only a number $N_s^2$ of reservoir simulations.

The parameters $N_{sim}$, $N_s^1$, $N_s^2$ and $N_{n,max}$ are not meant to be in the users' choice. Typical values are $N_{n,max} = 2\times N_r$, $N_{sim} = 2\times N_r$, $N_s^1 = 1$ and $N_s^2 = 1$. A users' choice is the maximum distance of selection for the neighborhood $d_{max}$, and which is a problem-dependent constant. An investigation of the impact of the choice of $d_{max}$ will be briefly shown in the next section through some examples.

---

[1] For each selected point $\mathbf{x}_i$ for the training set, we have a corresponding evaluation on a given realization. For the sake of notation simplicity we will denote the corresponding stored evaluation by $NPV(\mathbf{x}_i, R_i)$ although it is not necessarily evaluated on realization $R_i$.

An estimated standard deviation can also be included in the formulation of the estimated objective function $NPV^E$. In this case, the estimated objective function, which will not be tested in this paper, can be formulated as follows:

$$NPV^E(\mathbf{p}) = m^E(\mathbf{p}) + r\sigma^E(\mathbf{p}) \; , \qquad (14)$$

where $r$ is a constant, $m^E$ is defined as follows:

$$m^E(\mathbf{p}) = \frac{1}{S}\left[\sum_{i=1}^{N_r^2}\left(p_i NPV(\mathbf{p}, R_{j_i})\right) + \sum_{i=1}^{N_n}\left(\widetilde{p}_i NPV(\mathbf{x}_i, R_i)\right)\right] \; , \qquad (15)$$

and $\sigma^E$ is defined as follows:

$$\sigma^E(\mathbf{p}) = \sqrt{\frac{1}{S}\left[\sum_{i=1}^{N_r^2}\left(p_i\big(NPV(\mathbf{p}, R_{j_i}) - m^E(\mathbf{p})\big)^2\right) + \sum_{i=1}^{N_n}\left(\widetilde{p}_i\big(NPV(\mathbf{x}_i, R_i) - m^E(\mathbf{p})\big)^2\right)\right]} \; . \qquad (16)$$

## 4. Application of CMA-ES using the neighborhood approach on the PUNQ-S3 case

In this section, we apply the CMA-ES using the neighborhood approach -that we will call in the sequel the "using the neighborhood" approach- on the well placement problem on the benchmark reservoir case PUNQ-S3 (Floris et al., 2001). The model grid contains 19 cells in the x-direction, 28 cells in the y-direction and 5 cells in the z-direction. The cell sizes are 180m in the x and y directions and 18m in the z-direction. The elevation and the geometry of the field are shown in **Fig. 1**.

We consider 20 geological realizations that will be again denoted by $(R_i)_{i=1,\ldots,20}$. Each realization defines one possible porosity map and one possible permeability map. In these examples, the number of realization $N_r$ is then equal to 20.

We plan to drill two wells: one unilateral injector and one unilateral producer. The dimension of the problem is then equal to 12(= 6×2) corresponding to the number of parameters defining the coordinates of the wells to be drilled. The used parameterization is the same as defined by Bouzarkouna et al. (2011). In all the following applications, we use CMA-ES as an optimization algorithm with a population size equal to 40.

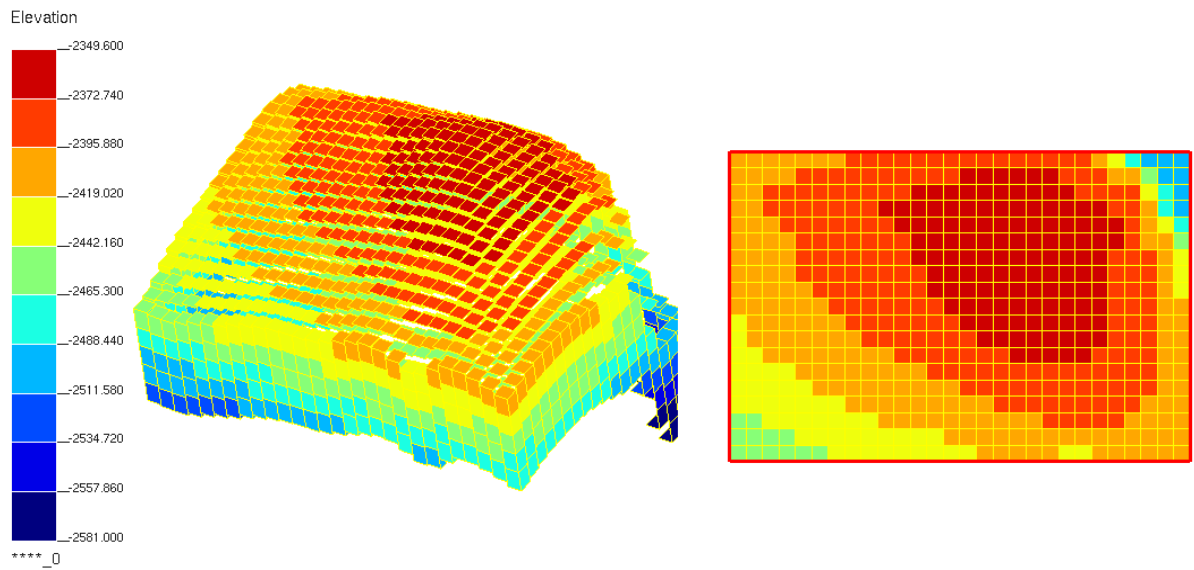

***Figure 1*** *Elevation (in meters) and geometry of the PUNQ-S3 test case.*

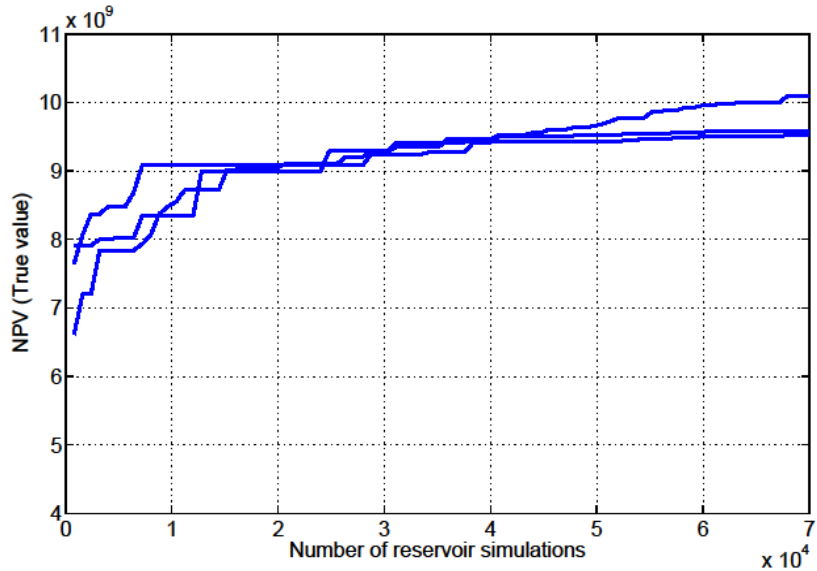

**Figure 2** *The evolution of the well placement optimization process on the PUNQ-S3 case using CMA-ES with the "using the mean of samples" approach. The best mean value of the NPV over the 20 possible realizations, i.e., NPV$^R$ is shown. Three runs are performed.*

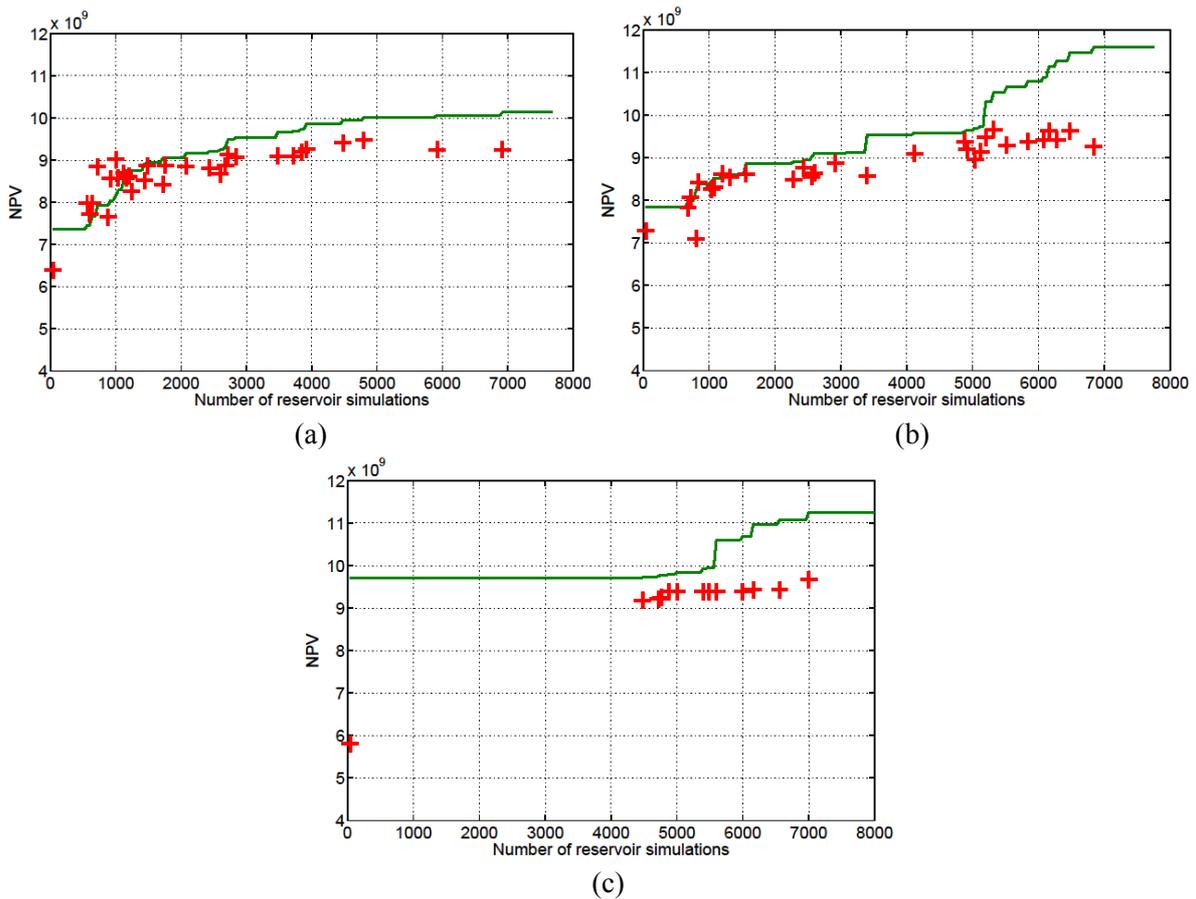

(a)

(b)

(c)

**Figure 3** *The evolution of the well placement optimization process on the PUNQ-S3 case using CMA-ES with the "using the neighborhood" approach, for three independent runs in (a), (b) and (c). The evolutions of the best estimated objective function, i.e., NPV$^E$ are drawn with green lines. The evaluations on the true objective function over the 20 possible realizations, i.e., NPV$^R$ are depicted with red crosses. The maximum distance of selection for the neighborhood $d_{\max}$ is equal to 4000.*

As a reference approach, we perform three independent runs in which we optimize the objective function $NPV^R$ as defined in Eq. (10). In this reference approach, we perform for each well configuration to be evaluated 20 reservoir simulations. The reference approach will be called in the sequel the "using the mean of samples" approach. **Fig. 2** shows the evolution of the best mean value of $NPV^R$, i.e., the NPV over the 20 possible realizations, for the three performed runs. The "using the mean of samples" approach is shown to be able to reach a mean value of $NPV^R$ equal to $9 \times 10^9$ using 15200 reservoir simulations. It is able also to reach a mean value of $NPV^R$ equal to $9.3 \times 10^9$ using 31200 reservoir simulations and a mean value of $NPV^R$ equal to $9.5 \times 10^9$ using 44400 reservoir simulations.

To evaluate the "using the neighborhood" approach, we use typical values of the parameters $N_{sim}$, $N_s^1$, $N_s^2$ and $N_{n,max}$ as defined in the previous section, i.e., $N_{n,max} = 2 \times N_r$, $N_{sim} = 2 \times N_r$, $N_s^1 = 1$ and $N_s^2 = 1$. We begin by choosing the maximum distance of selection for the neighborhood $d_{max}$ equal to 4000.

**Fig. 3** shows the evolution of the optimization process for three independent runs of CMA-ES with the "using the neighborhood" approach. The evolutions of the best estimated objective function, i.e., $NPV^E$ are drawn with green lines. During the optimization process, each own overall best point found on $NPV^E$, is evaluated on $NPV^R$. The evaluations performed on $NPV^R$ are depicted with red crosses. **Fig. 3** shows that when optimizing $NPV^E$, we are able to propose good points according to $NPV^R$ (points with an $NPV^R$ greater than $9 \times 10^9$). Moreover, $NPV^R$ tends to increase with an increasing number of performed reservoir simulations.

**Fig. 3(c)** shows a particular run in which the best $NPV^E$ value found at the first generation is equal to $9.7 \times 10^9$. This value is calculated according to Eq. (12), and thus calculated using only one single reservoir simulation (with one single random realization). Indeed, with a single reservoir simulation to evaluate one point, the estimated objective function cannot in general propose a good point according to $NPV^R$. Consequently, the best point found at the first generation according to $NPV^E$ has a "bad" $NPV^R$ value equal to $5.8 \times 10^9$. Thus, the optimization process does not propose for 112 iterations a new overall best point to be evaluated on $NPV^R$. The performance of this run can be avoided either by evaluating more often points using $NPV^R$ [2] or simply by using more reservoir simulations for each point to be evaluated at the beginning of the optimization, i.e., choosing $N_s^1 \geq 2$.

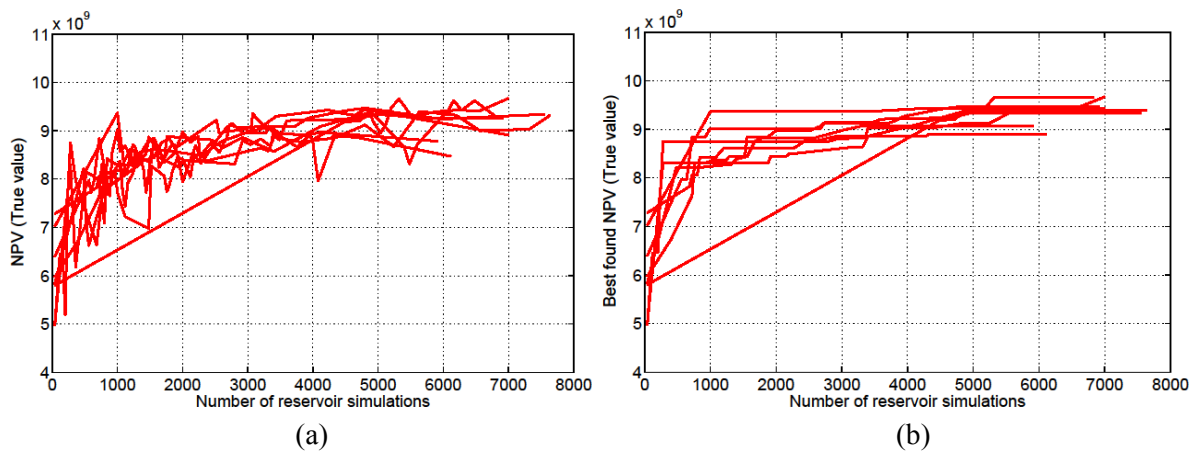

(a)                                                    (b)

***Figure 4*** *The evolution of the well placement optimization process on the PUNQ-S3 case using CMA-ES with the "using the neighborhood" approach for eight independent runs. (a) shows the evolution of the evaluations on* $NPV^R$. *(b) shows the evolution of the best found evaluation on* $NPV^R$. *The maximum distance of selection for the neighborhood* $d_{max}$ *is equal to* 4000.

---

[2] For example, one can evaluate the best found point according to $NPV^E$ at each iteration on $NPV^R$

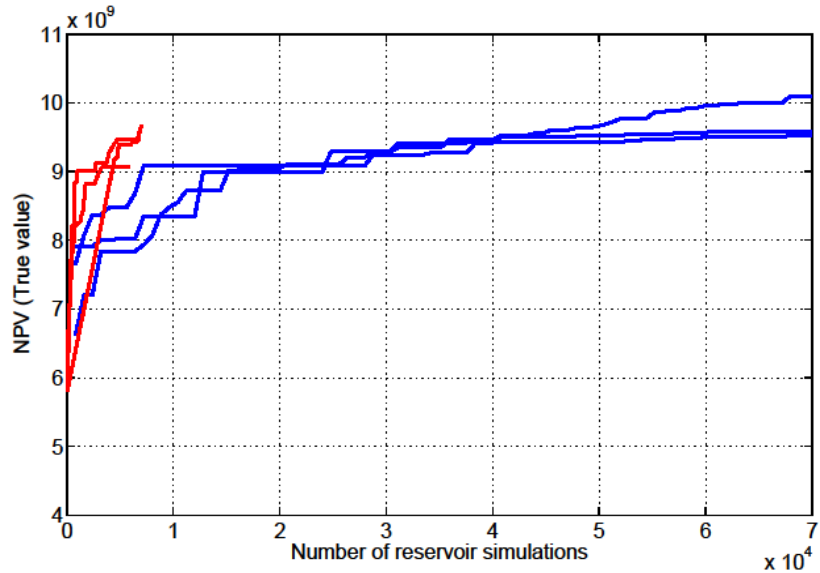

**Figure 5** *The evolution of the well placement optimization process on the PUNQ-S3 case using CMA-ES with the "using the mean of samples" approach and the "using the neighborhood" approach. The evolution of the best found evaluation on NPV$^R$ for the "using the neighborhood" approach is drawn with red lines. The evolution of the best found evaluation on NPV$^R$ for the "using the mean of samples" approach is drawn with blue lines. Three independent runs are performed for each approach. For the "using the neighborhood" approach, the maximum distance of selection for the neighborhood $d_{\max}$ is equal to 4000.*

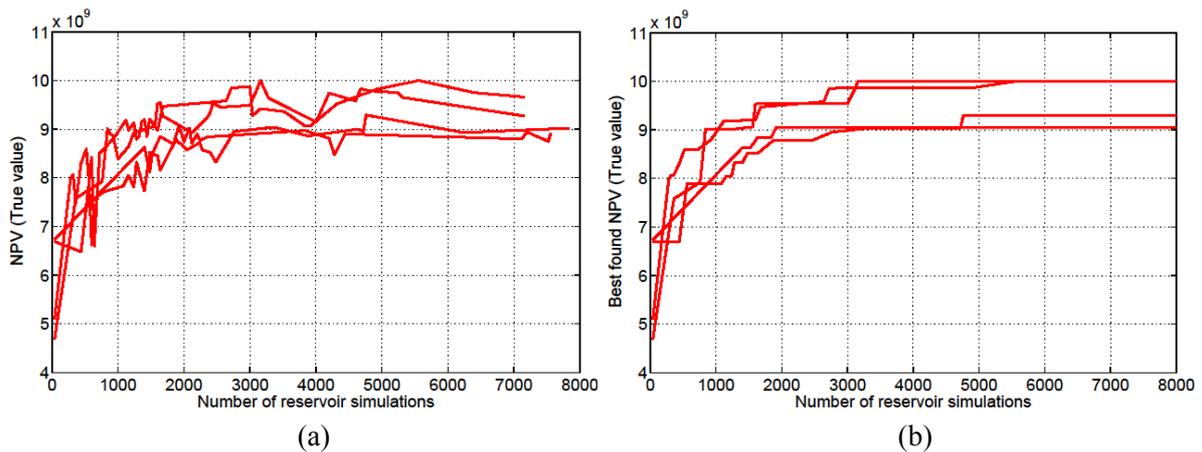

|     |     |
| :-: | :-: |
| (a) | (b) |

**Figure 6** *The evolution of the well placement optimization process on the PUNQ-S3 case using CMA-ES with the "using the neighborhood" approach for four independent runs. (a) shows the evolution of the evaluations on NPV$^R$. (b) shows the evolution of the best found evaluation on NPV$^R$. The maximum distance of selection for the neighborhood $d_{\max}$ is equal to 3000.*

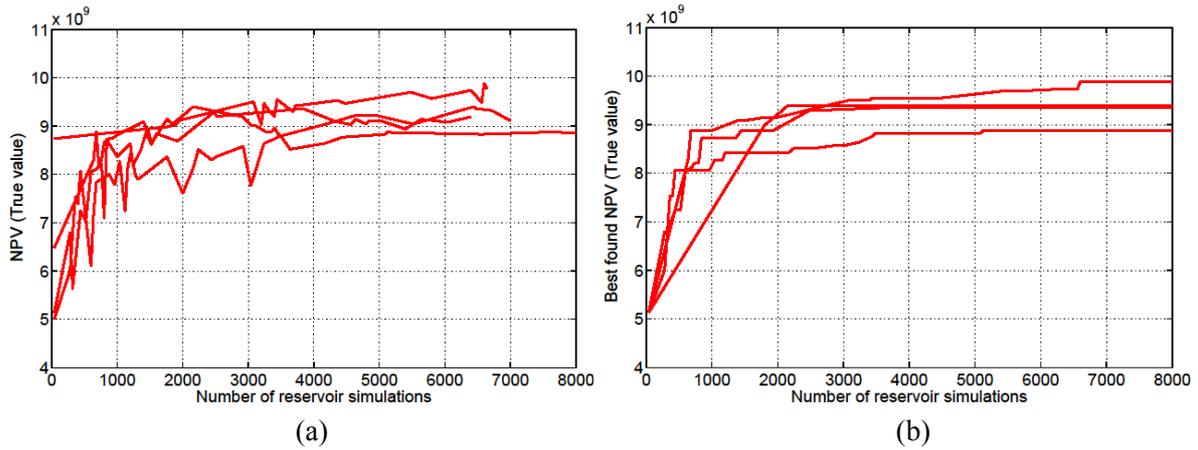

*(a)*        *(b)*

**Figure 7** *The evolution of the well placement optimization process on the PUNQ-S3 case using CMA-ES with the "using the neighborhood" approach for four independent runs. (a) shows the evolution of the evaluations on* $\text{NPV}^R$*. (b) shows the evolution of the best found evaluation on* $\text{NPV}^R$*. The maximum distance of selection for the neighborhood* $d_{\max}$ *is equal to 6000.*

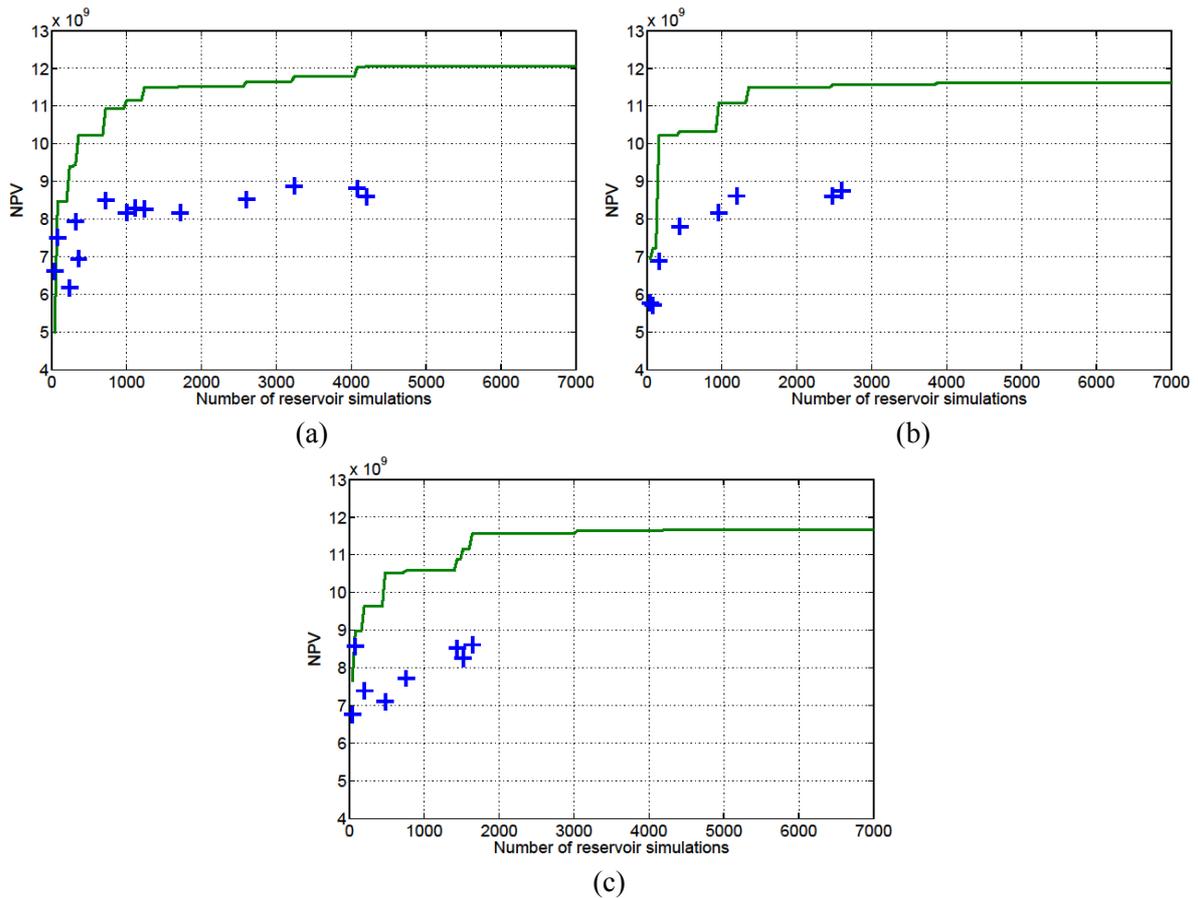

*(a)*        *(b)*

*(c)*

**Figure 8** *The evolution of the well placement optimization process on the PUNQ-S3 case using CMA-ES with the "using one realization" approach, for three independent runs in (a), (b) and (c). The evolutions of the best estimated objective function (equal to an evaluation on a randomly chosen realization) are drawn with green lines. The evaluations on the true objective function over the* 20 *possible realizations, i.e.,* $\text{NPV}^R$ *are depicted with blue crosses.*

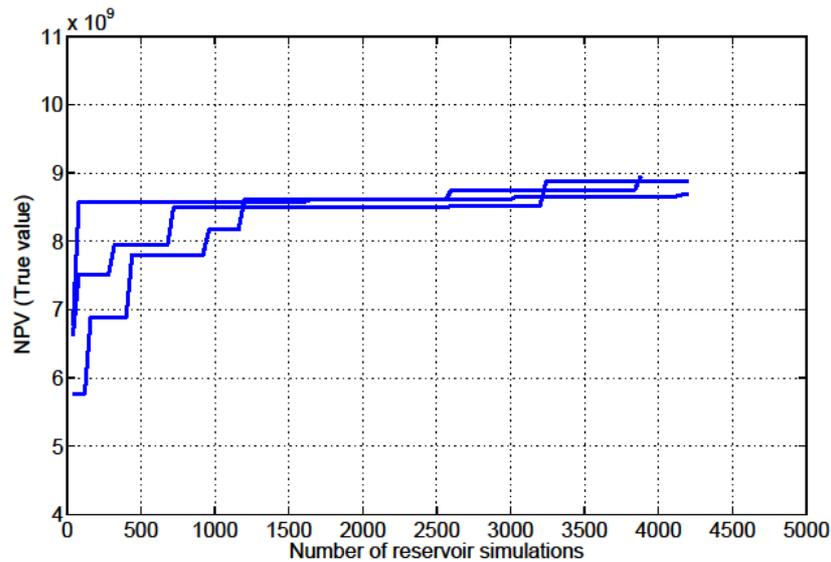

*Figure 9 The evolution of the well placement optimization process on the PUNQ-S3 case using CMA-ES with the "using one realization" approach. The best mean value of the* NPV *over the* 20 *possible realizations, i.e.,* $NPV^R$ *is shown. Three runs are performed.*

We show in **Fig. 4** the performance of eight independent runs of CMA-ES with the "using the neighborhood" approach. **Fig. 4(a)** shows the evolution of the evaluations performed on $NPV^R$. The evaluated points correspond to the best overall points found during the optimization process of $NPV^E$. **Fig. 4(b)** shows the evolution of the best evaluation performed on $NPV^R$. Seven runs out of the eight performed runs (88%) are able to reach an $NPV^R$ value greater than to $9 \times 10^9$, using a mean number of reservoir simulations equal to 2851. Consequently the reduction of the number of reservoir simulations to reach an $NPV^R$ greater than to $9 \times 10^9$ when using the "using the neighborhood" approach compared to the "using the mean of samples" approach is equal to 81%. Six runs out of eight performed runs (75%) are able to reach a value of $NPV^R$ greater than to $9.3 \times 10^9$, using a mean number of reservoir simulations equal to 4307, which offers a reduction of the number of reservoir simulations when comparing to the "using the mean of samples" approach equal to 86%. However, only two runs out of the eight performed runs (25%) are able to reach a value of $NPV^R$ greater than to $9.5 \times 10^9$. The mean number of reservoir simulations required to reach this value is 6160. Consequently the reduction of the number of reservoir simulations to reach an $NPV^R$ greater than to $9.5 \times 10^9$ when comparing to the "using the mean of samples" approach is again equal to 86%.

Three runs of CMA-ES with the "using the neighborhood" approach are shown together with the three performed runs of CMA-ES with the "using the mean of samples" approach in **Fig. 5**. Results show that although the "using the neighborhood" approach does not guarantee finding the best values of $NPV^R$ found by the "using the mean of samples" approach when comparing with the "using the mean of samples" approach, the number of reservoir simulations is reduced significantly by more than 81%.

The impact of the choice of the maximum distance of selection for the neighborhood $d_{max}$ is shown in **Figs. 6** and **7**. Comparing the results in Figs. **6**, **4** and **7** (with $d_{max} = 3000$, 4000 and 6000) shows that the approach is not very sensitive to the choice of $d_{max}$.

In the sequel, we compare the "using the neighborhood" approach with another approach in which the estimated objective function to be optimized is equal to an evaluation on a randomly chosen realization. This approach is called the "using one realization" approach. In this approach, we also evaluate on $NPV^R$ only the overall new best points found on the estimated objective function. **Figs. 8** and **9** show the evolution of the optimization process for three independent runs of CMA-ES with the "using one realization" approach. In **Fig. 8**, the evolutions of the best estimated objective function are again drawn with green lines. When comparing the "using the neighborhood" and the "using one realization" approaches through **Figs. 3** and **8**, it is clear that contrary to the "using the neighborhood" approach which is shown to be able to capture the geological uncertainty, the "using one realization" approach is shown to be not able to propose good points with high $NPV^R$. The three performed runs with the "using one realization" approach are not able to reach an $NPV^R$ value greater than $\$9 \times 10^9$.

## 5. Summary and discussions

In this paper, we have defined a new optimization approach under geological uncertainty with a reduced number of reservoir simulations. The approach uses the objective function evaluations of already simulated well configurations in the neighborhood of each well configuration. The proposed approach can be combined with any optimization algorithm. In this work, we show an application using the state-of-the-art stochastic optimizer CMA-ES on the well placement problem.

On the benchmark reservoir case PUNQ-S3, the proposed approach is shown to be able to capture the geological uncertainty while using only a reduced number of reservoir simulations. More specifically, the proposed approach is able to reduce significantly the number of reservoir simulations by more than 80% compared to the reference approach, i.e., the approach using all the possible realizations for each well configuration.

This work can be extended and enhanced by numerous means, mainly by using an adaptive strategy to define the parameters of the algorithm, although we suspect that the used values of the parameters defining the number of performed reservoir simulations can be a good choice.

The definition of the distance metric has a major influence on the neighbourhood selection, and thus on the definition of the objective function. Another improvement can then be achieved by using a Mahalanobis distance with respect to $\sigma^2 \mathbf{C}$ (instead of $\mathbf{C}$).

## Acknowledgements

The authors would like to thank Nikolaus Hansen and Christian Igel for their insightful comments on this work.